\documentclass[aps,prl,showpacs,twocolumn,groupedaddress,floatfix]{revtex4-1}
\usepackage{amssymb,amsmath}
\usepackage{graphicx}
\usepackage{subfigure}
\usepackage[english]{babel}
\usepackage{float}
\usepackage{color}

\newcommand{\be}{\begin{equation}}
\newcommand{\ee}{\end{equation}}

\begin{document}

\title{Localized modes in nonlinear photonic kagome nanoribbons}

\author{Mario I. Molina}

\affiliation{Departamento de F\'{\i}sica, MSI-Nucleus on Advanced Optics,  and Center for Optics and Photonics (CEFOP), 
Facultad de Ciencias, Universidad de Chile, Santiago, Chile.}

%\date{\today}

\begin{abstract}
We examine localization of light in nonlinear (Kerr) kagome lattices in the shape of narrow  strips of varying width. For the narrowest ribbon, the band structure features a flat band leading to linear dynamical trapping of an initially localized excitation. We also find a geometry-induced bistability of the nonlinear modes as the width of the strip is changed. A crossover from one to two dimensions localization behavior is observed as the width is increased, attaining two-dimensional behavior  for relatively narrow ribbons.
\end{abstract}

%\ocis{190.4420; 190.6135; 190.1450; 190.4350}

\maketitle

\clearpage

The kagome lattice has persisted as an object of interest in several branches of physics since its appearance in 1951\cite{1951}. In condensed matter physics, kagome lattices are valued because
they lead to spin frustration when the system contains antiferromagnetic interactions. Many properties are still a matter of speculation, however,  such as the nature of its magnetic ground state\cite{sachdev}. 
In ultracold atom physics, several optical lattices have been used so far, most of which are primitive Bravais lattices. More recently, however, recent works have used non-standard optical lattices that contain a few-sites basis, such as the kagome lattice, hoping that the low energy degrees of freedom arising from the placing of a particle inside the unit cell, would bring richer dynamics and ordering\cite{ultracold}. Very recently, a magnetic field induced metal- insulator transition has been predicted in kagome nanoribbons\cite{dey}.   
In photonics, kagome lattices have been recently proposed as a cladding in hollow-core polymer fibers, where it has been observed that the low overlap of the core modes and the microstructured material, plus the low density of states in the cladding, can lead to an improved core guidance\cite{fiber}. The existence and stability of gap solitons and vortices in infinite, nonlinear kagome lattices have been examined in ref.\cite{kevrekidis}. In this work we focus on finite, ribbon-like kagome lattices and study the onset of localized modes and their stability as the width of the ribbon is changed, going from a quasi one-dimensional geometry to a quasi two-dimensional one.

Let us consider a two-dimensional kagome photonic lattice in the form of a long ribbon, i.e., a $N\times M$ lattice of width $w=M\ll N$ (Fig.1). This type of waveguide array can be easily manufactured by means of the direct femtosecond laser-writing technology\cite{szameit}. Another choice is the use of optical induction\cite{cornelia}.
In the framework of the coupled-mode theory, the electric field
$E({\bf r})$ propagating along the waveguides can be presented as
a superposition of the waveguide modes, 
\be
E({\bf
r})=\sum_{\bf n} E_{\bf n} \phi({\bf r}-{\bf n}),\label{eq:1}
\ee
where
$E_{\bf n}$ is the amplitude (in units of (Watt)$^{1/2}$) of the (single) guide mode
$\phi({\bf r})$ centered on the site with 
lattice number ${\bf n}= (n_1,n_2)$.  The evolution equations
for the modal amplitudes $E_{\bf n}$ are:
\be i {d E_{\bf n}\over{dz}} +  \sum_{\bf m} V_{{\bf n},{\bf m}}
E_{\bf m} + \gamma |E_{\bf n}|^2 E_{\bf n} =
0,\label{eq:2}
\ee
where ${\bf n}$ denotes the position of a guide center, $z$ is the longitudinal 
distance (in meters), $V_{{\bf n},{\bf m}}
$ is the coupling between nearest-neighbor guides
(in units of 1/meter) and $\gamma$ is the nonlinear
coefficient (in units of 1/(Watt $\times$ meter)), defined by $\gamma=\omega_{0} n_{2}/c A_{\mbox{eff}}$, where 
$\omega_{0}$ is the angular frequency of light, $n_{2}$ is the nonlinear coefficient
of the guide and $A_{\mbox{eff}}$ is the effective area of the linear modes. The nonlinear parameter $\gamma$ is normalized to $1$
for the focussing nonlinearity.

Next, we analyze the stationary localized modes of Eq.~(\ref{eq:1})
in the form $E_{\bf n}(z) = E_{\bf n} \exp(i \beta z)$, where the
amplitudes $E_{\bf n}$ satisfy the nonlinear difference equations,
\be %
 -\beta E_{\bf n} + \sum_{\bf m} V_{{\bf n},{\bf m}} E_{\bf m} + \gamma |E_{\bf
n}|^2 E_{\bf n} = 0
\label{eq:2}
\ee
%%%%%%%%%%%%%%%%%%%%%%%%%%%%%%%%%%%%%%%
\begin{figure}[t]
\begin{center}
\noindent
\includegraphics[scale=0.75]{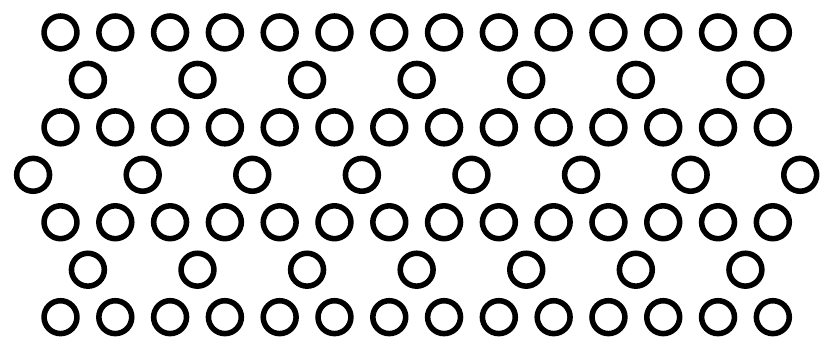}
\caption{Schematics of an optical kagome lattice nanoribbon of width $w=3$. The ribbons are of  indefinite extension in the horizontal direction, remaining finite in the vertical direction.}
\label{fig1}
\end{center}
\end{figure}
%%%%%%%%%%%%%%%%%%%%%%%%%%%%%%%%%%%%%%%%

For a given value of $\beta$ the system is solved by a multi-dimensional Newton-Raphson method, using the anticontinuum (i.e., decoupled) limit as an initial condition. For the fundamental mode, we look for a localized solution with maximum at the center of the array and decaying along and across the ribbon. In order to visualize the modes, we will use a generic gaussian mode $\phi(x,y) = \exp[-(x^2+y^2)/\sigma]$ in Eq.(\ref{eq:1}), with $\sigma=0.15$. as we will see, the results depend substantially upon the ribbon width. Each mode family is characterized by its power content $P=\sum_{n} |E_{n}|^2$ as a function of the propagation constant $\beta$.

The linear stability of each nonlinear mode is computed in the usual manner by defining a weakly perturbed mode as $E_{n}(z)=(E_{n}+u_{n}(z)+i v_{n}(z)) \exp(i \beta z)$, where $u_{n}(z)$ and $v_{n}(z)$ are real. Their evolution equations can be expressed in compact form by defining real vectors $\delta {\bf U}=\{u_{n}\}$ and $\delta {\bf V}=\{v_{n}\}$ and real matrices ${\bf A}=\{A_{n m}\}=\{ \delta_{n,m+1}+\delta_{n,m-1}+(-\beta+3\gamma |E_{n}|^{2})\delta_{n,m}$ and 
${\bf B}=\{B_{n m}\}=\{ \delta_{n,m+1}+\delta_{n,m-1}+(-\beta+\gamma |E_{n}|^{2})\delta_{n,m}$. The dynamical evolution of the perturbation can then be expressed as $\delta \ddot{{\bf U}} + {\bf B}{\bf A}\ \delta{\bf U}=0$ and $\delta \ddot{{\bf V}} + {\bf A}{\bf B}\ \delta{\bf V}=0$, where an overdot denotes a derivative in $z$. Thus, the linear stability of the mode depends upon the eigenvalue spectra of ${\bf A}{\bf B}$ and ${\bf B}{\bf A}$: If any of the real eigenvalues is negative, the mode is unstable; otherwise the solution is stable.

\noindent
{\em Bulk modes}: The narrowest ribbon has width $w=1$ and consists of a chain of hexagonal rings with 
a basic cell containing $5$ sites (Fig.2). The linear ($\gamma=0$) dispersion relation can be computed in closed form and consists of five bands: $\beta=2$, $\beta=\pm \sqrt{2 + 2 \cos(k)}$, $\beta=1 \pm \sqrt{3+2 \cos(k)}$. The whole spectrum is gapless and one of the bands is completely flat, just as in the infinite lattice case. The existence of the flat band has implications for the diffusion of an initially localized excitation:
If the initial profile has significant overlap with the modes on the flat band, then there will be a portion of the excitation that cannot propagate away from the initial site due to the vanishing  group velocity of the band. As a consequence, there will be {\em linear} selftrapping at long evolution times. This localization effect is due to a 
destructive interference effect due to the inherent geometry of the lattice. All this is borne out very nicely by a long-time computer simulation of the propagation of an initially localized pulse placed in the middle of the ribbon. Since the ribbon is really finite in the `long' direction, absorbing boundary conditions
were used to avoid reflections from each end. The results are shown in Fig.\ref{trapping} which shows  the intensity of the field vs the evolution distance for the initial site and several other sites close and far from it. We see a localized  optical profile around the initial guide, as predicted.

For the nonlinear regime ($\gamma\neq 0$), Fig. 4 shows the power vs propagation constant curve for the 1D fundamental mode. It reaches all the way to the band edge, as is typical for 1D lattices.
%%%%%%%%%%%%%%%%%%%%%%%%%%%%
\begin{figure}[h]
\centering
\includegraphics[width=0.35\textwidth]{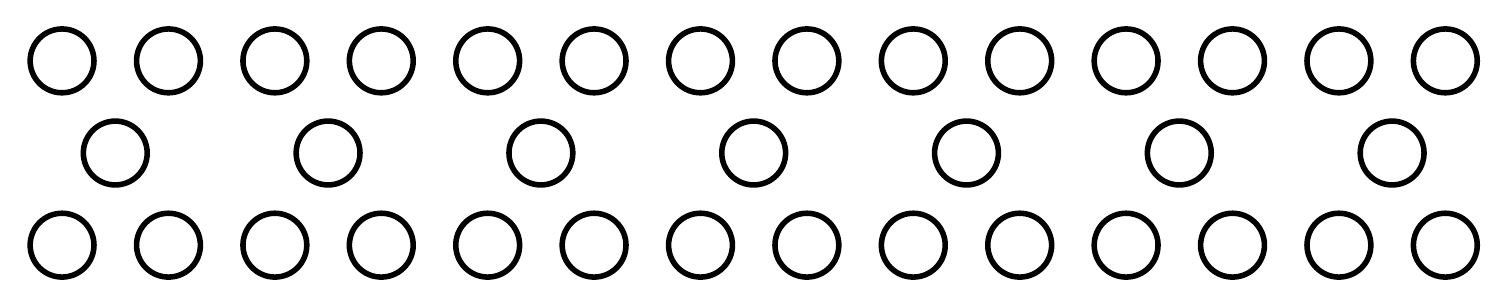}
\vspace{0.5cm}

\includegraphics[width=0.35\textwidth]{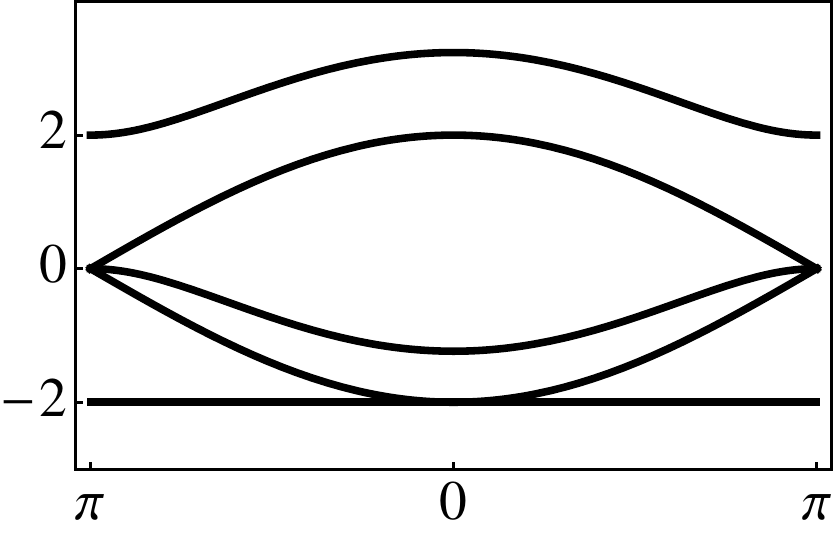}
\caption{Top: 1D Kagom\'{e} Ribbon. Bottom: Linear bands for a 1D Kagom\'{e} Ribbon.}

\label{fig2}
\end{figure}
%%%%%%%%%%%%%%%%%%%%%%%%%%%%
%%%%%%%%%%%%%%%%%%%%%%%%%%%%
\begin{figure}[h]
\centering
\includegraphics[width=0.5\textwidth]{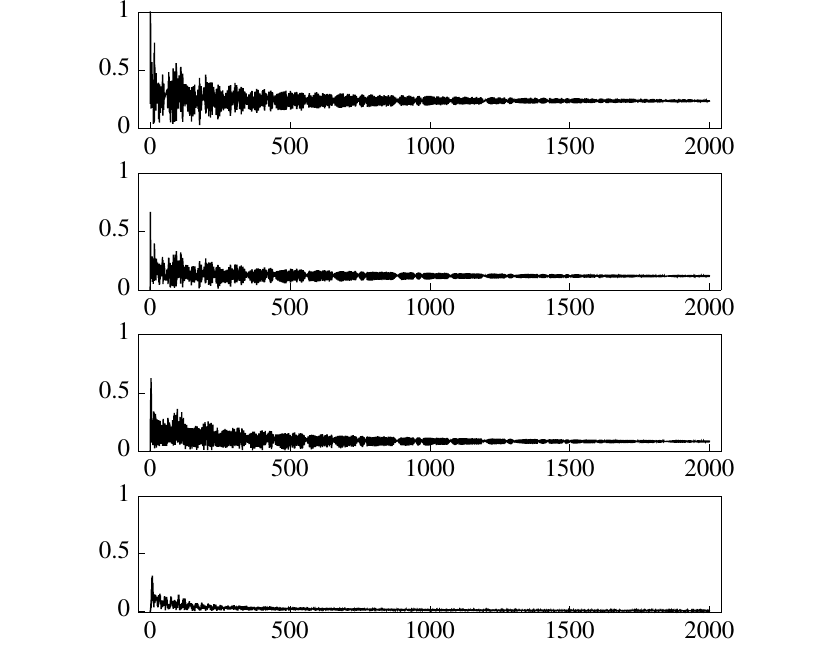}
\vspace{0.5cm}
\caption{Dynamical evolution of initially localized excitation. The plot shows the amplitude of the field as a function of longitudinal distance (`time'), at several sites away from the initial one. From top to bottom: At initial site, at a first nearest-neighbor, at a third nearest-neighbor and  at an eighth nearest-neighbor.  }
\label{trapping}
\end{figure}
%%%%%%%%%%%%%%%%%%%%%%%%%%%%

%%%%%%%%%%%%%%%%%%%%%%%%%%%%%%%%%%%%%%%%%%%%
\begin{figure}[h]
\centering
\includegraphics[width=0.25\textwidth]{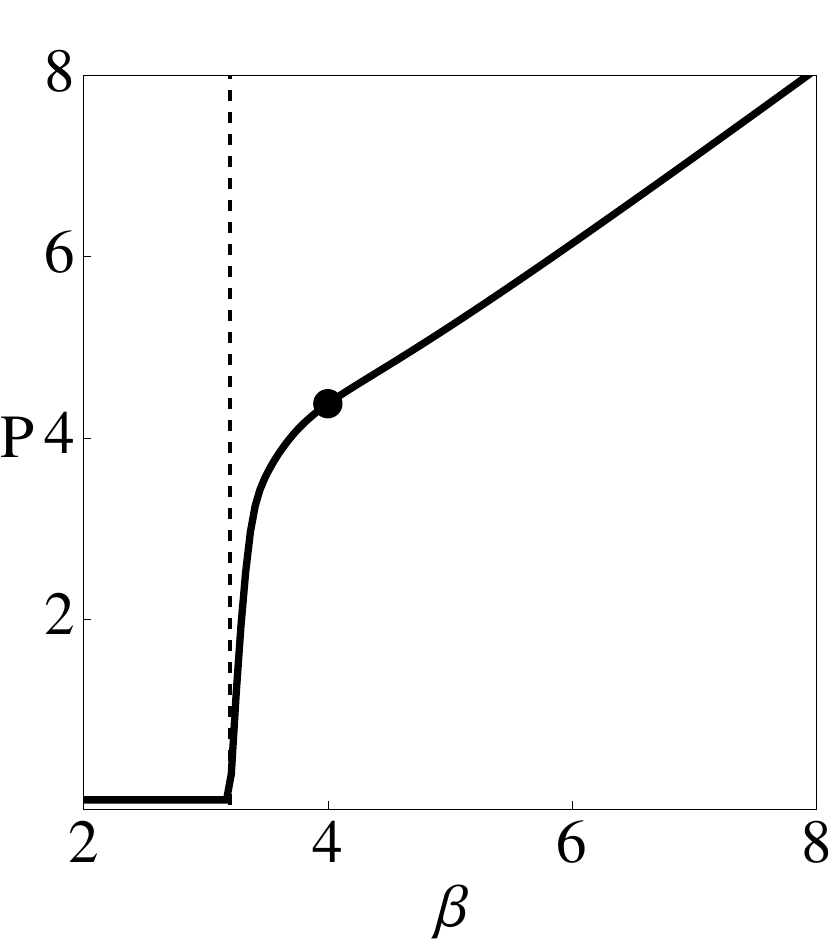}
\hspace{0.7cm}
\raisebox{0.4cm}
{\includegraphics[width=0.075\textwidth]{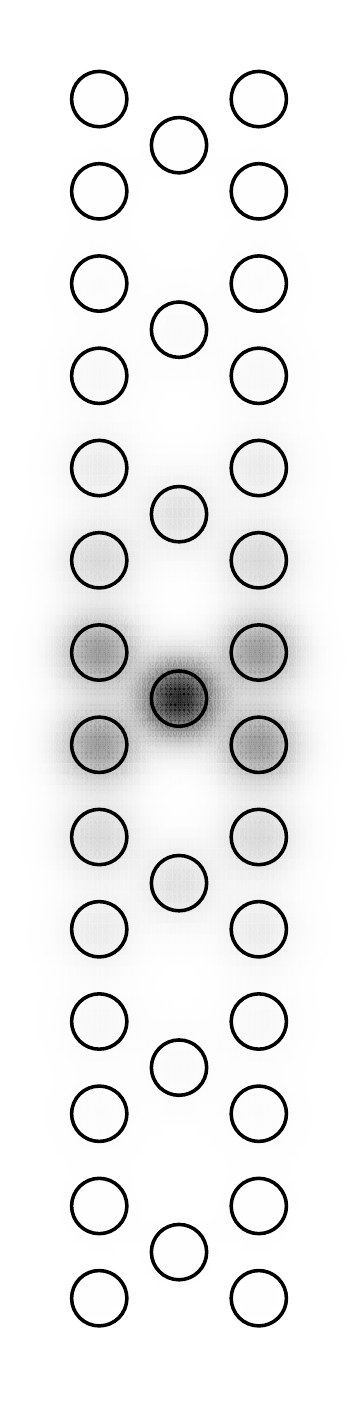}}
\caption{Left: Power vs propagation constant curve for the fundamental odd mode for a width $w=1$ ribbon (``one-dimensional''). Vertical line marks the position of the band edge. Right: Intensity distribution for fundamental single-site mode marked with black circle.   Shading is proportional to the intensity distribution of the mode power.}
\label{fig3}
\end{figure}
%%%%%%%%%%%%%%%%%%%%%%%%%%%%

We now increase the ribbon width to $w=2$ and compute again the power vs propagation constant curve (Fig.5). This time the curve features a stable and an unstable regime, and does not reach the band edge. This is a two-dimensional-like behavior, which is surprising given the narrowness of the ribbon. For $w=3$ (Fig.6), we see still a different behavior: Now the ribbon behaves like a finite 2D lattice, characterized by a bistable power vs propagation constant curve. On each side of the bistable curve we have a stable mode whose width depends on its closeness to the band: It is wider (narrower) for the mode whose propagation constant is farther (closer) to the band edge. Separating these two stable modes, there is an unstable one.

For $w=4$ and higher (Fig.7), we are back to the two-dimensional -like behavior, with a power vs propagation constant curve that is stable and decreases with decreasing propagation constant until its slope vanishes, followed by a unstable regime with negative slope, which is in agreement with the Vahitov-Kolokolov stability criterium for the fundamental mode\cite{VK}.

%%%%%%%%%%%%%%%%%%%%%%%%%%%%%%%%%%%%%%%%%%%%
\begin{figure}[h]
\centering
\includegraphics[width=0.25\textwidth]{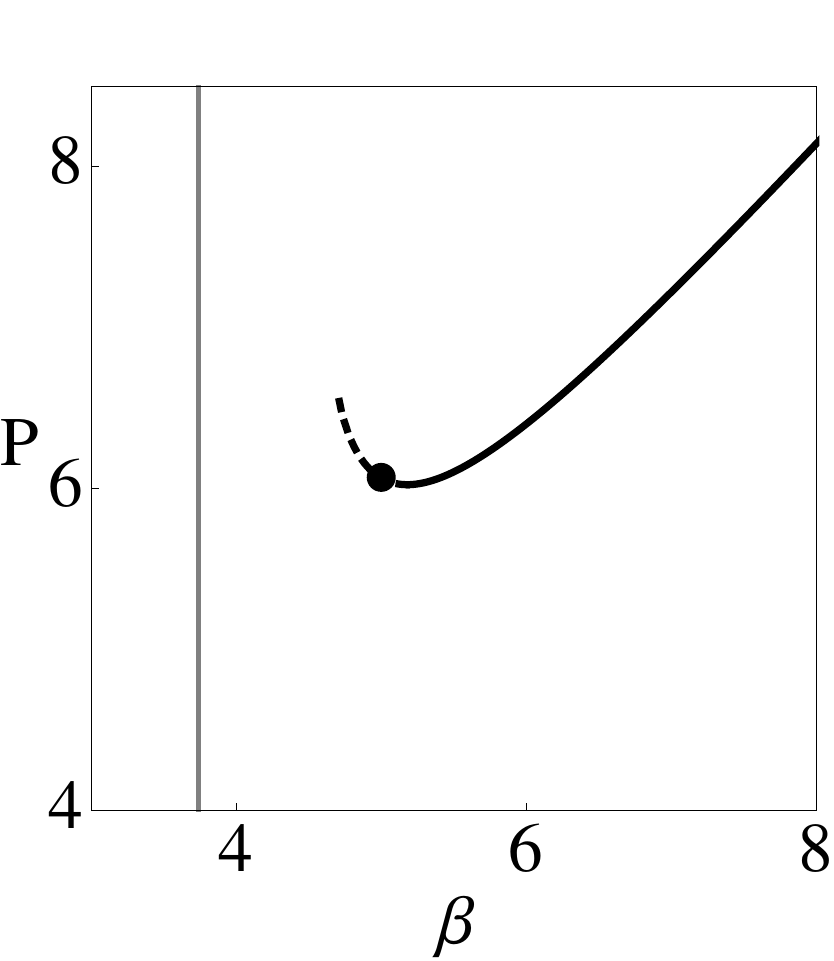}
\hspace{0.7cm}
\raisebox{0.25cm}
{\includegraphics[width=0.1\textwidth]{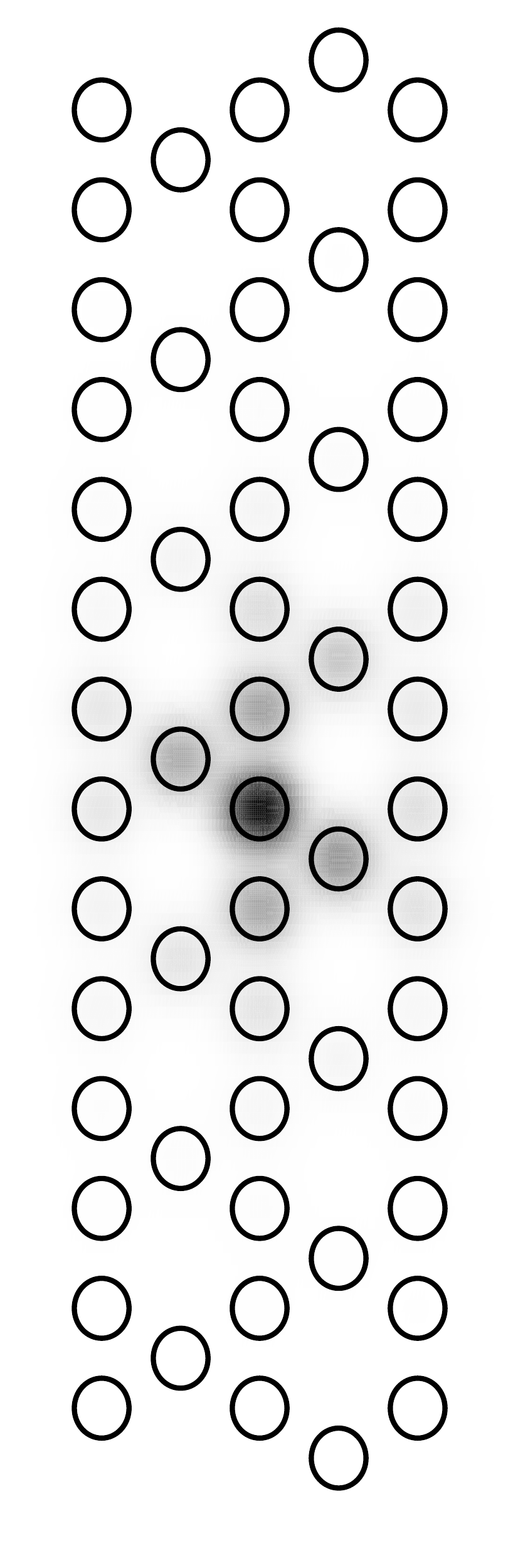}}
\caption{Left: Power vs propagation constant curve for the fundamental odd mode for a width $w=2$ ribbon. Solid (dashed) portion denotes stable (unstable) regime. Vertical line marks the position of the band edge. Right: Intensity distribution for fundamental single-site mode marked with black circle.   Shading is proportional to the intensity distribution of the mode power.}
\label{fig3}
\end{figure}
%%%%%%%%%%%%%%%%%%%%%%%%%%%%

%%%%%%%%%%%%%%%%%%%%%%%%%%%%%%%%
\begin{figure}[h]
\centering
\includegraphics[width=0.25\textwidth]{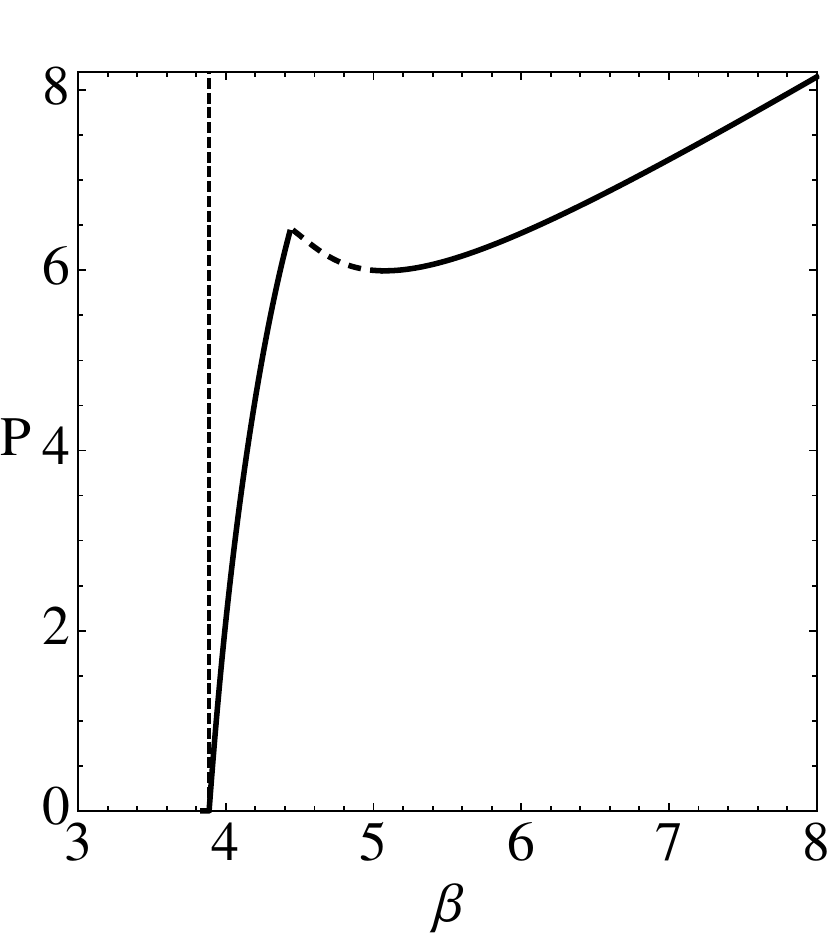}
\hspace{0.7cm}
\raisebox{0.25cm}
{\includegraphics[width=0.125\textwidth]{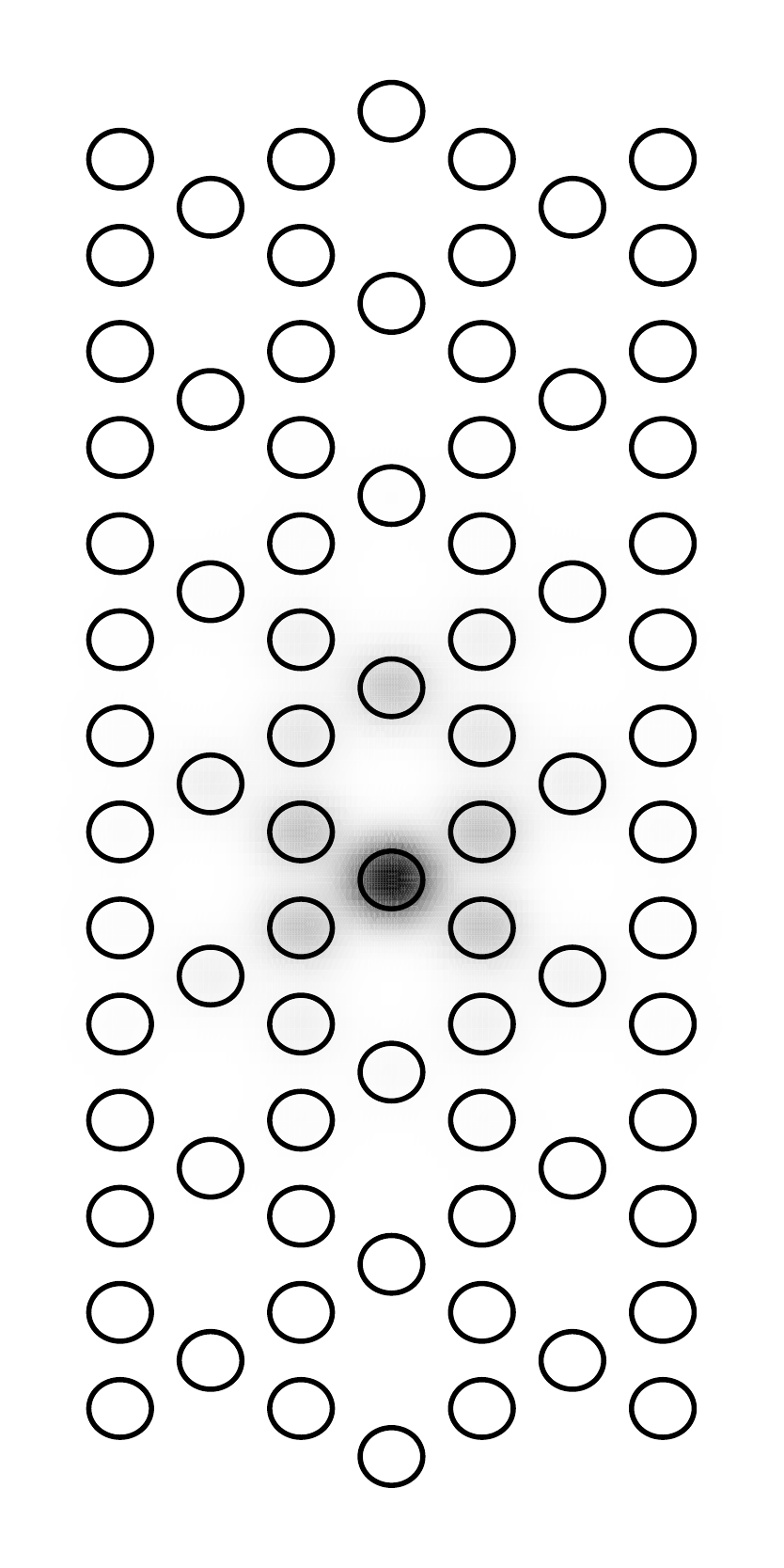}}
\caption{Left: Power vs propagation constant curve for the fundamental odd mode of  width $w=3$ ribbon. Solid (dashed) portion denotes stable (unstable) regime. Vertical line marks the position of the band edge. Right: Intensity distribution for fundamental single-site mode marked with black circle.   Shading is proportional to the intensity distribution of the mode power.}
\label{fig5}
\end{figure}
%%%%%%%%%%%%%%%%%%%%%%%%%%%%

\begin{figure}[h]
\centering
\includegraphics[width=0.225\textwidth]{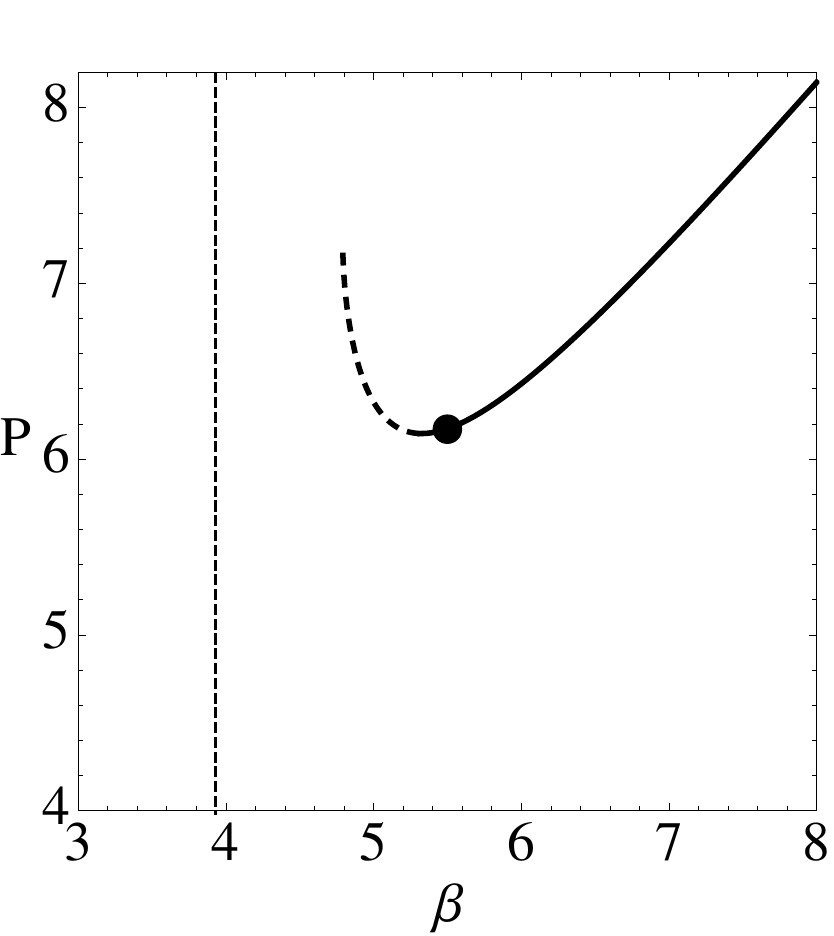}
\hspace{-0.1cm}
\raisebox{0.25cm}
{\includegraphics[width=0.25\textwidth]{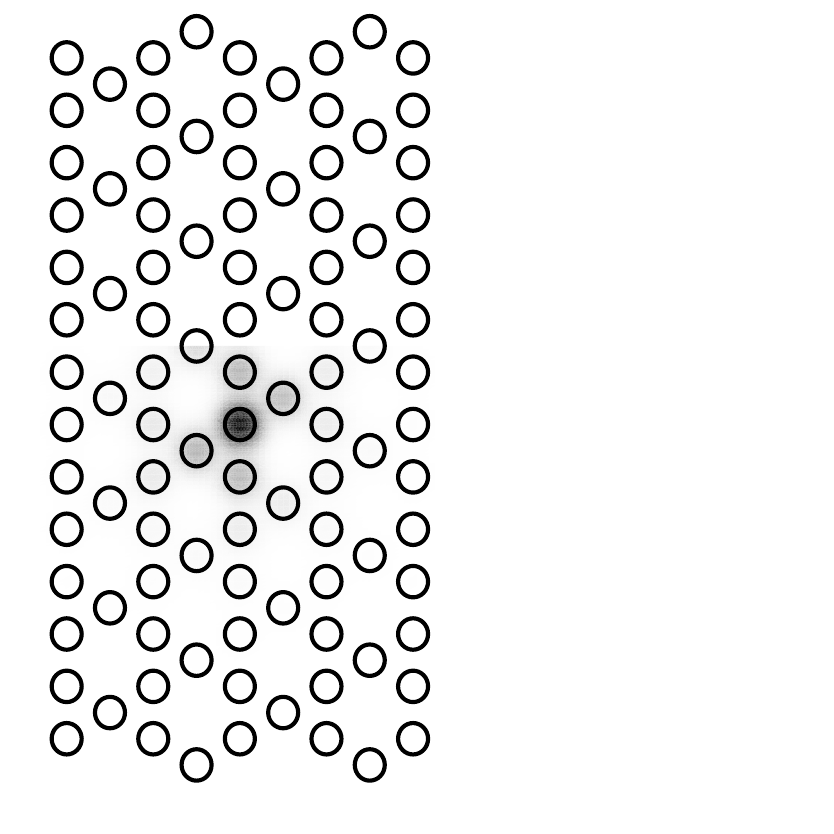}}
\caption{Left: Power vs propagation constant curve for the fundamental odd mode of a width $w=4$ ribbon. Solid (dashed) portion denotes stable (unstable) regime. Vertical line marks the position of the band edge. Right: Intensity distribution for fundamental single-site mode marked with black circle.   Shading is proportional to the intensity distribution of the mode power.}
\label{fig6}
\end{figure}
%%%%%%%%%%%%%%%%%%%%%%%%%%%%

{\em Surface modes}: We compute nonlinear surface modes centered on a single waveguide at the very edge. A finite ribbon sample features two different
types of edge: Straight one along the `long' direction and `zig-zag' along the short direction (Fig.1).  

For a given ribbon width $w$,  the power vs propagation constant curve for modes located on the straight edge exhibits the usual behavior already observed in other two-dimensional lattices, like the 
existence of a power threshold to generate the mode. This is observed for all $w$ values examined, ranging from $w=1$ up to $w=5$. For widths greater than $4$, the corresponding $P$ vs $\beta$ curves  
 become indistinguishable from each other on the scale shown here. On the `zig-zag' boundary there are three non-equivalent surface sites to consider, characterized by their different coordination numbers (2,3 and 4). Their power vs propagation constant curves are qualitatively similar. An example of such curves for $w=3$ is shown in Fig.8. The lowest threshold power in this case corresponds to the mode centered on the zig-zag boundary with $2$ nearest-neighbors.

Thus, we have examined localization properties of two-dimensional photonic lattices consisting of kagome nanoribbons. For the quasi-one-dimensional case, and in the absence of nonlinearity,  we found a flat band and the ensuing linear localization of dynamical excitations, which agrees with the phenomenology found earlier for infinite kagome lattices. As the width of the ribbons increased, we found a crossover between one-and two-dimensional soliton behavior, characterized by the appearance of a bistable regime, which constitutes an example of a geometry-induced bistability for solitons. This type of bistability could be generic of two-dimensional finite lattices, as evidenced by similar results obtained for the graphene lattice\cite{graphene}.
%%%%%%%%%%%%%%%%%%%%%%%%%%%%%%%%%%%%%%%%%%%%%%%%%
\begin{figure}[h]
\centering
\includegraphics[width=0.23\textwidth]{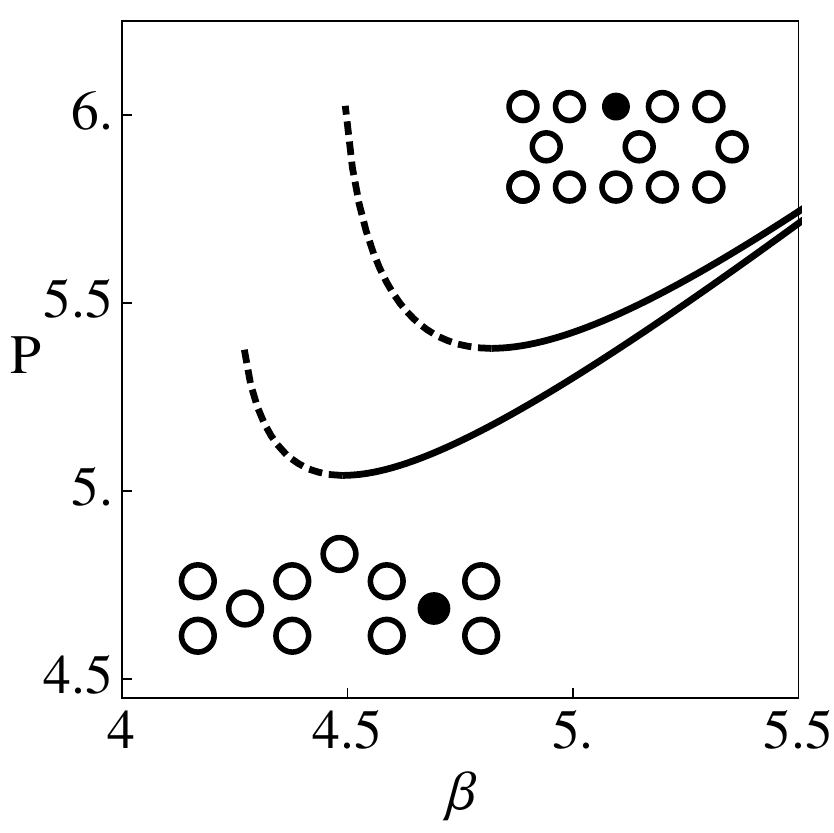}
\includegraphics[width=0.23\textwidth]{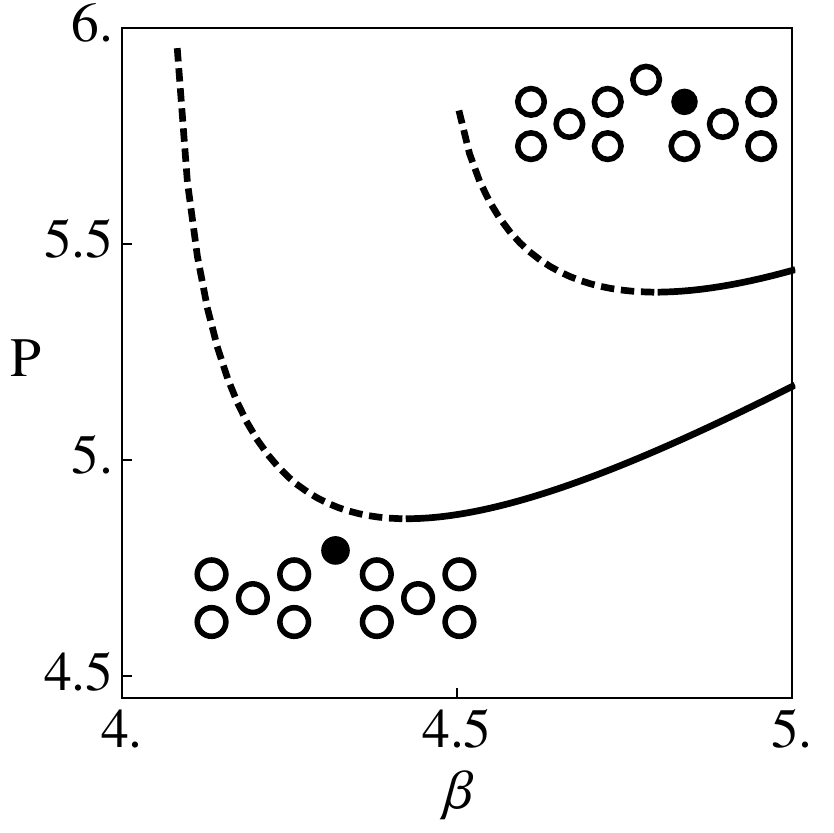}
\caption{Left: Power vs propagation constant curve for several surface modes of a width $w=3$ ribbon. Solid (dashed) portion denotes stable (unstable) regime. The insets show the vicinity of the array boundary and the the black circle marks the position of the mode center. }
\label{fig7}
\end{figure}
%%%%%%%%%%%%%%%%%%%%%%%%%%%%%%%%%%%%%%%%%%%%%%%%%%%
The author is grateful to Y. S. Kivshar and P. G. Kevrekidis for useful discussions.  This work was supported in part by Fondo Nacional de Ciencia y Tecnolog\'{\i}a (Grants 1080374 and 1120123), Programa Iniciativa Cient\'{\i}fica Milenio (Grant P10-030-F), and Programa de Financiamiento Basal (Grant FB0824/2008).

\end{document}